\documentclass[aps,prb,twocolumn,amsmath,amssymb,superscriptaddress,floatfix]{revtex4-1}
\usepackage[utf8]{inputenc}
\usepackage{textcomp}
\usepackage{amsmath}
\usepackage{amssymb}
\usepackage{graphicx,graphics}
\usepackage{latexsym,verbatim}
\usepackage{dcolumn} 
\usepackage{color}
\usepackage{cancel}
\usepackage[breaklinks]{hyperref}

\usepackage{graphicx}  
\usepackage{dcolumn}   
\usepackage{bm}        

\usepackage{graphics}
\usepackage{latexsym}
\usepackage{amsmath}
\usepackage{float}
\usepackage{pstricks,pst-plot}

\begin{document}

\title{Signatures of surface Majorana modes in the magnetic response of topological superconductors}

\author{Luca Chirolli}
\affiliation{IMDEA-Nanoscience, Calle de Faraday 9, E-28049 Madrid, Spain}

\author{Francisco Guinea}

\affiliation{IMDEA-Nanoscience, Calle de Faraday 9, E-28049 Madrid, Spain}
\affiliation{School of Physics and Astronomy, University of Manchester, Manchester, M13 9PY, UK}

\begin{abstract}
We study the magnetic susceptibility of a two-dimensional cone of Majorana modes localised at the surface of a 
three-dimensional time-reversal invariant topological superconductor belonging to class DIII. A field parallel to the 
surface tilts the surface Majorana cone along the supercurrent direction. For fields  larger than a critical threshold 
field $H^*$, $H>H^*$, a transition from type I to type II Dirac cone occurs and a finite current carried by the Majorana 
modes start to flow, leading to an additional diamagnetic contribution to the surface magnetization. On a curved surface 
interband transitions are promoted by the Majorana spin connection that couples to the external field, giving rise to a 
finite frequency magnetic susceptibility.   
\end{abstract}

\maketitle

\section{Introduction}

Topological Superconductors (TSCs) are a very promising state of matter in which a topological  superconducting gap opens in the bulk of a system and leads to the confinement at the surface of unconventional Andreev states named Majorana states  \cite{Majorana1937,Kitaev2001,Alicea}. Majorana modes constitute a class of topologically protected surface excitation appearing at the boundary of topological states of matter \cite{QiRMP2011,AndoFu-review,Bernevig-book,HasanRMP2010,Chiu,Altland,SchneyderPRB2008,Kitaev}, and they represent one of the basic resources in topological quantum computation \cite{Ivanov2001,TewariZoller2007,nayak}.  Recently, doped Bi$_2$Se$_3$ topological insulators (TI) \cite{FuPRL2007,ZhangNatPhys2009,Fang,ChenScience2009} have been suggested as candidates that may realise odd-parity, time-reversal invariant (TRI) topological superconductivity \cite{FuBerg,HorPRL2010,KrienerPRL2011,peng-prb2013,SasakiPRL2011,wang,wray-np2010,venderbos2015} belonging to class DIII. In three dimensions these systems are expected to host Majorana modes forming a Dirac cone in the basis of Majorana Kramers partners\cite{ZhangKaneMele,Ryu,Volovik-book,Qi-PRL2009,Qi-PRB2010}. The localised and charge neutral character of Majorana modes has induced most of the theoretical detection proposals and the experimental efforts to focus on local spectroscopy, Josephson effect, and interferometry, to find a proof of their existence \cite{Beenakker,Mourik1003,Ramon}. As far as class DIII topological superconductors is concerned, the presence of surface Majorana modes is expected to produce a strongly anisotropic spin susceptibility \cite{ZhangPRL2009}.  

Recently, the authors have shown that for class DIII topological superconductors a coupling to a vector potential can arise at finite momentum and finite energy \cite{Chirolli-SpinConn2018}.  In a planar geometry a magnetic field laying on the surface of the system produces a supercurrent, that Doppler shifts the Majorana modes and results in a tilting of the cone. The tilting direction in momentum space is orthogonal to the applied field and parallel to the supercurrent. When Majorana modes are confined on a curved surface an additional coupling of geometric origin arises, that involves the Majorana spin connection \cite{Chirolli-SpinConn2018}. The latter generates finite matrix elements between empty and occupied states of the surface Majorana cone and a response is expected at finite frequency. Majorana modes can be detected through the application of time-dependent orbital magnetic fields. 

\begin{figure}[b]
\includegraphics[width=0.45\textwidth]{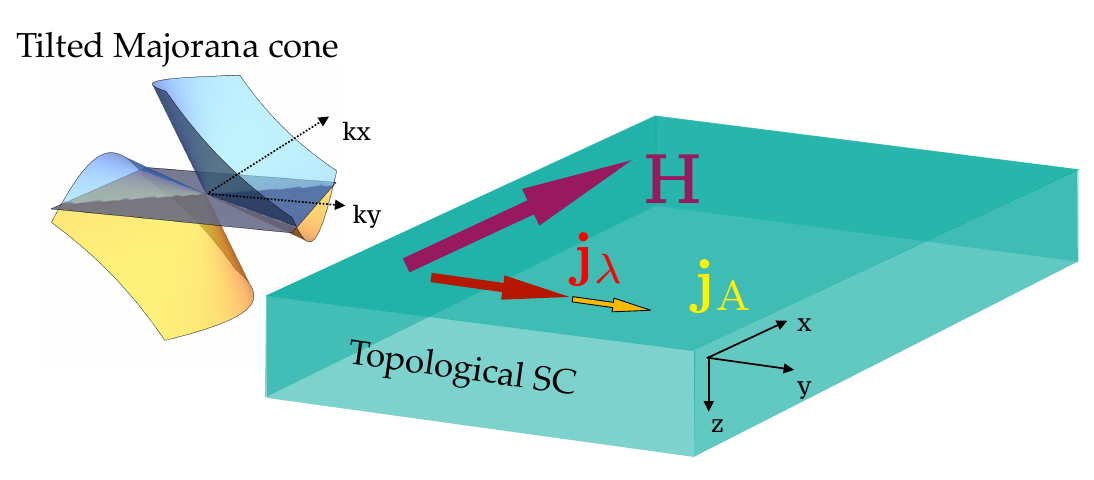}
\caption{Schematics: an external magnetic field ${\bf H}$ is forced to lie on the surface of the system by Meissner screening. The associated supercurrent ${\bf j}_\lambda$ Doppler shifts the Majorana cone, resulting in a cone tilting along the ${\bf j}_\lambda$ direction. For $H>H^*$ a transitions to type II overtilted cone occurs and the supercurrent acquires a component ${\bf j}_{\rm A}$ carried by the Majorana modes.  \label{Figure1}}
\end{figure}

In this work we study the magnetic susceptibility and the frequency dependent response of a Majorana cone localised at the surface of a class DIII three-dimensional topological superconductor. The Majorana velocity determines a threshold field $H^*$  for which a structural change of the Majorana cone takes place, that is characterised by a transition from type I to type II cone. For $H>H^*$, in the overtilted regime, a finite Andreev current flows carried by Majorana modes, as schematically depicted in Fig.~\ref{Figure1}. The Andreev current adds to the supercurrent and participates to Meissner screening of the external field by generating an additional surface diamagnetic magnetization. For small-band-gap doped Dirac insulators characterised by odd-parity superconductivity, such as the $A_{1u}$ phase predicted in Bi$_2$Se$_3$ \cite{FuBerg}, the Majorana velocity affecting $H^*$ can be tuned by changing the chemical potential, in a way that the value of $H^*$ falls into the Meissner phase for strong doping. Signatures of Majorana modes are then expected to appear in the magnetic susceptibility, with a signal amplitude scaling as the ratio $\lambda/L_z$ between the penetration depth $\lambda$ and the sample thickness $L_z$. In type II superconductors characterised by an appreciable ratio $\lambda/L_z$ the additional signal becomes detectable. Additionally, in systems characterised by a finite surface curvature, we find that the emergence of the Majorana spin connection in the tilting term, together with a curvature-induced non-zero Zeeman term, gives rise to a finite frequency magnetic susceptibility in response to a time-dependent magnetic field. Our findings acquire a universal character in finite geometries and open the way to detection of Majorana states via thermodynamic bulk measurements, in contrast with the widely used local spectroscopy probes.

The work is structured as follows: In Sec.~\ref{Sec:system} we introduce the system under study, in Sec.~\ref{Sec:coupling} we describe the coupling to an external magnetic field, in Sec.~\ref{Sec:andreev} we calculate the Andreev current for a planar geometry and in Sec.~\ref{Sec:orb-susc} we calculate the associated additional orbital magnetic susceptibility due to the surface Majorana modes. In Sec.~\ref{Sec:curvature} we consider a spherical system and give the surface Hamiltonian, and in Sec.~\ref{Sec:freq-resp} we study the finite frequency response. In Sec.~\ref{Sec:conclusions} we conclude the work with final remarks.

\section{The system}
\label{Sec:system}

We start our analysis by considering a specific example of an odd-parity superconductor realised in doped TI, such as the one proposed for doped Bi$_2$Se$_3$ [\onlinecite{FuBerg}]. The mean field Hamiltonian in the Nambu basis $\psi_{\bf k}=({\bf c}_{\bf k},is_y{\bf c}^\dag_{-{\bf k}})^T$, with $c_{{\bf k},s,i}$ a fermionic state of momentum ${\bf k}$, spin $s$, and $p_z$-like orbital $i=T,B$ for the top/bottom layers in the $k\cdot p$ approximation, reads ($\hbar=1$)
\begin{equation}\label{HamTI}
{\cal H}=\tau_z(m\sigma_x+v\sigma_z(k_xs_y-k_ys_x)+v_zk_z\sigma_y-\mu)+\Delta \tau_x\sigma_ys_z,
\end{equation}
where $\sigma_i$ are Pauli matrices spanning the two-fold orbital space, $m$ is the insulating band-gap, $v$ the Dirac velocity, and $\Delta>0$ the mean-field value of the superconducting order parameter. The Hamiltonian Eq.~(\ref{HamTI}) is time-reversal invariant, with ${\cal T}=is_y\hat{K}$ and $\hat{K}$ complex conjugation, centrosymmetric, with ${\cal P}=\sigma_x$ the parity operator, and it has a full bulk gap of size $\Delta$ on the Fermi surface. The system realises a TSC belonging to class DIII and hosts a surface Majorana cone localised at the boundary of the system. 

We consider a semi-infinite system of doped Bi$_2$Se$_3$ occupying the $z>0$ region. The realistic boundary condition compatible with the quintuple layer structure of the crystal is $\sigma_z\psi(z=0)=-\psi(z=0)$ (for the $z<0$ region the realistic boundary condition is $\sigma_z\psi(z=0)=\psi(z=0)$ [\onlinecite{HsiehPRL2012}]).  The system realises a TI if the condition ${\rm sign}(mv_z)<0$ is satisfied. Additionally, for finite $\Delta>0$ a zero-energy Majorana Kramers pair is found in the region $z>0$ at $k_x=k_y=0$ with wavefunction $\psi_\alpha(z)=|\alpha\rangle\phi(z)$  [\onlinecite{HsiehPRL2012}], with
\begin{equation}\label{Eq:MKP}
\phi(z)=e^{-z/\xi_z}\left[
\begin{array}{c}
\sin(k_Fz)\\
{\rm sign}(m)\sin(k_Fz-\theta)
\end{array}\right]_\sigma,
\end{equation}
where $k_F=\sqrt{\mu^2-m^2}/|v_z|$, $e^{i\theta}=(|m|+i|v_z|k_F)/\mu$, $\xi_z=|v_z|/\Delta$ is the SC coherence length along the $z$-direction. The states $|\alpha\rangle=[(1,-\alpha)_s,-i~{\rm sign}(v_z)(1,\alpha)_s]_\tau$, with $\alpha=\pm 1$, are simultaneous eigenstates of the mirror helicity $\tilde{M}=-is_x\tau_z$ [\onlinecite{HsiehPRL2012}], with eigenvalues $i\alpha$, and of the operator $\tau_ys_z$, $\tau_ys_z|\alpha\rangle=-{\rm sign}(v_z)|\alpha\rangle$. For definiteness we choose $m<0$ and $v_z>0$. Projecting the Hamiltonian Eq.~(\ref{HamTI}) onto the subspace spanned by the states $|\psi_\alpha\rangle$ we find the surface Hamiltonian describing a Majorana cone 
\begin{equation}
{\cal H}_{\bf k}=v_\Delta(k_x\alpha_y+k_y\alpha_z),
\end{equation}
where $\alpha_i$ are Pauli matrices in the basis $|\psi_\alpha\rangle$ and the velocity of the Majorana modes is $v_\Delta\simeq v|m|\Delta/\mu^2$ (for a TI the sign of $v_\Delta$ is opposite to the sign of the TI surface states, that for $m<0$ have negative velocity $-v$) [\onlinecite{HsiehPRL2012}]. It follows that the strength and sign of the velocity $v_\Delta$ depend on the topological character of the doped insulator.

\section{Coupling to the magnetic field}
\label{Sec:coupling}

The presence of an external magnetic field is accounted for by a minimal coupling substitution in the Nambu basis ${\bf k}\to {\bf k}+e{\bf A}({\bf r})\tau_z$. For weak field we neglect the spatial dependence of the gap and the resulting coupling Hamiltonian reads
\begin{eqnarray}\label{Eq:HA}
{\cal H}_B&=&{\cal H}_{\bf A}+\frac{1}{2}g\mu_B{\bf s}\cdot{\bf B},\\
{\cal H}_{\bf A}&=&ev\sigma_z(A_xs_y-A_ys_x)+ev_z\sigma_yA_z,\label{Eq:HA}
\end{eqnarray}
with ${\bf B}=\nabla\times{\bf A}$ the full induction field and $g$ the material $g$-factor. If we take the matrix elements of the orbital coupling on the states Eq.~(\ref{Eq:MKP}) we find $\langle\psi_\alpha|{\cal H}_{\bf A}|\psi_{\alpha'}\rangle=0$. A non-zero coupling occurs at finite momentum. We then split the Hamiltonian Eq.~(\ref{HamTI}) in three terms, ${\cal H}={\cal H}_0+{\cal H}_{\bf k}+{\cal H}_\Delta$, with ${\cal H}_0=\tau_z(m\sigma_x-\mu)$, ${\cal H}_\Delta=\Delta s_z\sigma_y\tau_x$. Defining the in-plane effective mass $\hat{M}^{-1}=2v^2(\mu-m\sigma_x)/(\mu^2-m^2)$, the effective orbital coupling at small momentum can be written as ${\cal H}_{{\bf k}\cdot{\bf A}}=-{\cal H}_{\bf A}{\cal H}_{0}^{-1}{\cal H}_{\bf k}+{\rm H.c.}$, that amounts to
\begin{eqnarray}\label{HpA}
{\cal H}_{{\bf p}\cdot{\bf A}}=\frac{e}{2}\hat{M}^{-1}\frac{v_i^2}{v^2}\{A_i,k_i\}+
\frac{e}{2}\hat{M}^{-1}\tilde{g}_iS_iB_i,
\end{eqnarray}
where $\{A_i,p_i\}=A_ik_i+k_iA_i$,  the anisotropic $g$-factor is $\tilde{g}_x=\tilde{g}_y=v_z/v$ and $\tilde{g}_z=1$, and we used a gauge for which $\nabla\cdot{\bf A}=0$. The spin operator ${\bf S}$ appearing in Eq.~(\ref{HpA}) is ${\bf S}=(\sigma_xs_x,\sigma_xs_y,s_z)$ and corresponds to the correct spin operator that behaves as a pseudo-vector under the point group $D_{3d}$ operations that characterise the Hamiltonian Eq.~(\ref{HamTI}) \cite{Chirolli-2017}.

When an external field is applied to the system, Meissner screening forces the external field to lay in the plane defined by the surface of the system. We assume the external field to be applied along the $x$ direction, ${\bf H}=H{\bf e}_x$, as schematically depicted in Fig.~\ref{Figure1}. In the Landau gauge the vector potential corresponding to the bulk screened field reads ${\bf A}=(0,H\lambda e^{-z/\lambda},0)$. For an in-plane field the matrix elements of the Zeeman coupling are zero on the Majorana Kramers pair Eq.~(\ref{Eq:MKP}). We are then only left with the orbital term, that has no structure in spin nor in Nambu space, and can only amount to a diagonal term in the Majorana subspace. The full Hamiltonian projected on the Majorana cone reads
\begin{equation}\label{HsurfA}
{\cal H}_{\bf k}=v_\Delta(k_x\alpha_y+k_y\alpha_z)+v_Hk_y,
\end{equation}
where the magnetic field dependent velocity $v_H$, in the limit $\lambda\gg\xi_z$, is given by 
\begin{equation}
v_H\equiv\langle\psi_\alpha|e\hat{M}^{-1}A_y(z)|\psi_\alpha\rangle=\lambda eH2v^2/\mu,
\end{equation} 
Defining $m^*=\mu/2v^2$, we can write the field dependent velocity as $v_H= eH\lambda/m^*$, in analogy with the results of Ref.~[\onlinecite{Chirolli-SpinConn2018}].  Introducing the coherence length $\xi=v/\Delta$ and equating the two velocities $v_H=v_\Delta$ we find the threshold tilting field (restoring $\hbar$)
\begin{equation}
H^*=\frac{\Phi_0}{2\pi\xi\lambda}\frac{|m|}{\mu}=\eta H_c,
\end{equation}
with $\eta=|m|/\mu$ and $\Phi_0=h/2e$. In the second equality we express the threshold field in terms of the thermodynamic critical field $H_c=\Phi_0/(2\pi \xi\lambda)$  [\onlinecite{DeGennes}]. By tuning $\eta$ the threshold field can be shifted in the Meissner phase.

\section{Andreev current}
\label{Sec:andreev}

The spectrum of the surface Majorana Hamiltonian has two branches, $\epsilon_{{\bf k},\pm}=\pm v_\Delta+v_Hk_y$. At $H=H^*$ one of the two branches, $\epsilon_{{\bf k},-}$, becomes flat along the line $k_x=0$ and a structural change of the dispersion takes place, that is characterised by a formally diverging number of states at zero energy \cite{Chirolli-SpinConn2018}. The current operator is obtained by the usual relation $\hat{j}_i=\partial {\cal H}_{\rm BdG}/\partial A_i$. At zero temperature the Andreev current density carried by the Majorana modes is given by 
\begin{equation}
j^A_y(z)=\frac{e}{m^*}|\phi(z)|^2\sum_\alpha\int \frac{d^2{\bf k}}{(2\pi)^2}k_y\Theta(-\epsilon_{{\bf k},\alpha}),
\end{equation} 
where the $\Theta$-function restricts the integral to the occupied states. For $H<H^*$ there are as many occupied states at positive $k_y$ as at negative $k_y$, so that the current is zero. On the other hand, an Andreev current will flow for $H>H^*$. The tilting takes below the Fermi level previously occupied states which lay within an angular sector $-\phi^*\leq \phi\leq \phi^*$, with $\phi^*=\cos^{-1}(H^*/H)$. Defining the surface quasiparticle density $\rho_{\rm 2D}=k_F^2/4\pi$ we find
\begin{equation}\label{Eq:AndCurr}
j^A_y(z)=-ev_F\frac{2\rho_{\rm 2D}}{\xi}e^{-2z/\xi}\sqrt{1-(H^*/H)^2},
\end{equation}
that predicts an abrupt diamagnetic signal as $H^*$ is approached. Formally, the derivative with respect to $H$ of $j_y^A(z)$ diverges at $H=H^*$, as result of the structural change of the  band structure, in which the zero energy level becomes macroscopically occupied.

\section{Orbital Magnetic Susceptibility}
\label{Sec:orb-susc}

In order to understand what is the effect of the Andreev current carried by the Majorana modes for $H>H^*$, we have to self-consistently solve for the vector potential. We assume Meissner screening acting locally in the bulk following the dependence $j_y^\lambda(z)=-A_y(z)/(4\pi \lambda^2)$, where $\lambda=\sqrt{m^*/(4\pi e^2 N_s)}$ is the penetration depth, with $N_s$ the number of bulk superconducting electrons. Majorana modes produce an additional contribution $j_y^A(z)\equiv j_y^A e^{-2z/\xi}$, that is restricted to the surface on a scale $\xi/2$ and adds to the bulk screening current $j_y^\lambda(z)$, so that the total current is $j_y(z)=j_y^\lambda(z)+j_y^A(z)$, as schematically depicted in Fig.~\ref{Figure1}. The total current is related to the vector potential by the Maxwell equation $4\pi j_y(z)=-\partial_z^2A_y(z)$. It follows that the self-consistent vector potential $A_y(z)$ necessarily acquires an additional component, $A_y(z)=A_0 e^{-z/\lambda}+A_1 e^{-2z/\xi}$. The coefficient $A_0$ and $A_1$ are obtained by imposing $h_x(z=0)=H$, with the induction field $h_x(z)=-\partial_zA_y(z)$, that produces the constraint 
\begin{align}
A_0=\lambda H-2\lambda A_1/\xi.
\label{constraint}
\end{align}
The amplitude of $j_y^A$ in Eq.~(\ref{Eq:AndCurr}) depends non-linearly on the vector potential at the surface $A_y$ via the velocity $v_H=e\langle A_y(0)\rangle/m^*=e(A_0+A_1/2)/m^*$ and  the problem amounts in solving the following non-linear equation 
\begin{equation}\label{EqA1}
A_1=4\pi\times \frac{1}{2}ev_F\rho_{\rm 2D}\xi \sqrt{1-\left(\frac{H^*\lambda}{A_0+A_1/2}\right)^2}.
\end{equation}

\begin{figure}[t]
\includegraphics[width=0.48\textwidth]{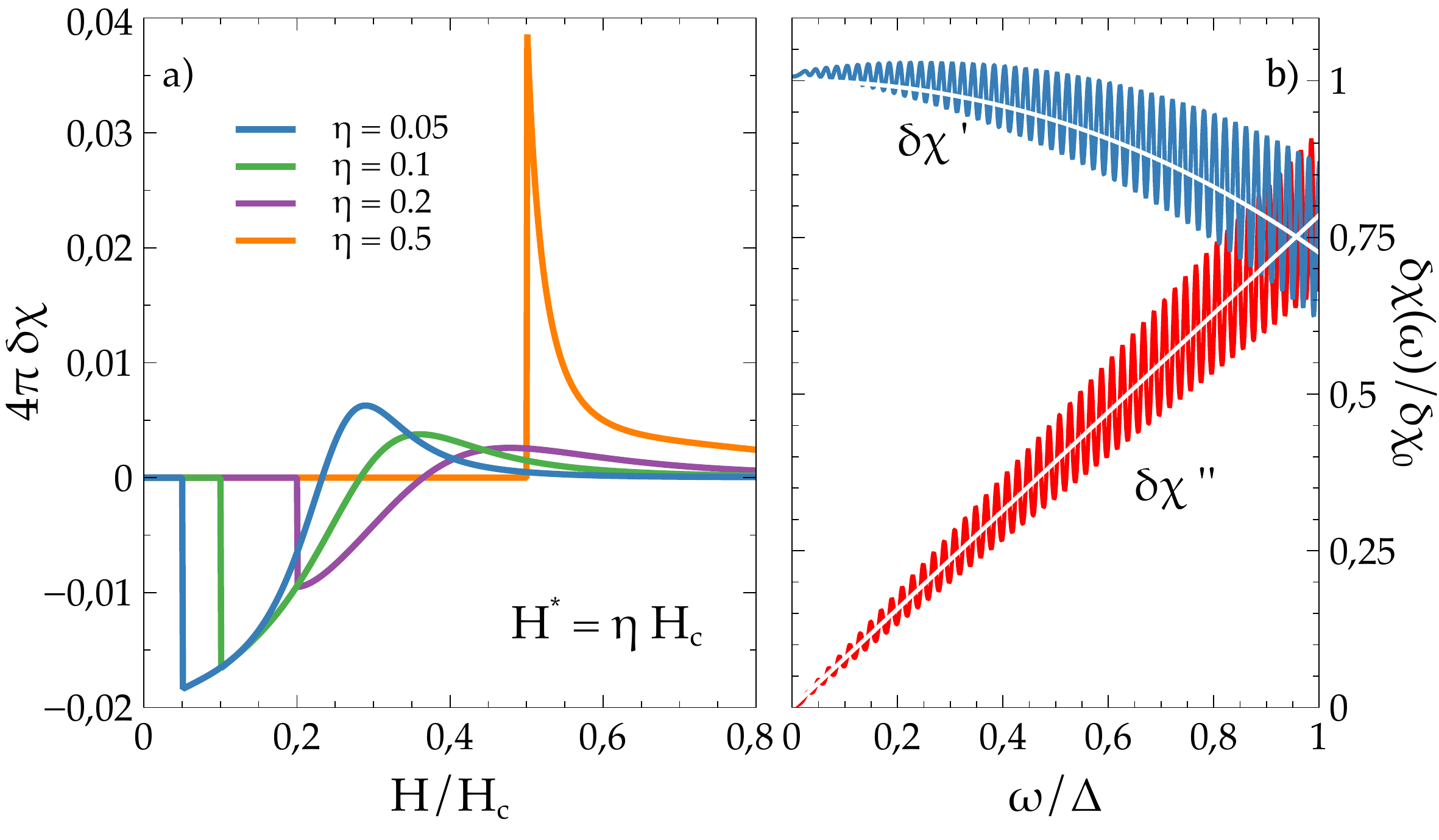}
\caption{a) Orbital susceptibility $\delta\chi$ as given by Eq.~(\ref{deltachi}) for different values of $\eta$. In the plot we chose $\epsilon=\epsilon_0/\eta$, with $\epsilon_0=0.2$, $\lambda/\xi=10$, $\lambda/L_z=0.02$, and we have expressed the external field in units of $H_c$. b) Real $\delta\chi'$ and imaginary $\delta\chi''$ part of the finite frequency orbital surface susceptibility $\delta\chi(\omega)$ in a spherical geometry versus the applied field for $k_FR=100$. Fast oscillations reflect the discrete nature of the spectrum with level spacing $\Delta/k_FR$. White lines show the $k_FR\to\infty$ asympthotics as given by Eq.~(\ref{Eq:EnvChi}). \label{Figure}}
\end{figure}

The magnetization of the system is calculated through the Gibbs free energy of the system \cite{DeGennes}, $G=\int d{\bf r}({\bf h}^2-4\pi{\bf j}\cdot{\bf A}-2{\bf H}\cdot{\bf h})/(8\pi SL_z)=-HA_y(0)/(8\pi L_z)$, where $S$ is the surface area and $L_z$ the thickness of the system. The magnetization is found as $M=(B-H)/4\pi$, where the induction field is $B=-4\pi\partial G/\partial H$. For $H<H^*$ we find $M=M_0\equiv -H(1-\lambda/L_z)/4\pi$, that agrees with the thermodynamic limit $M=-H/4\pi$ for $L_z\to \infty$. For $H>H^*$ the magnetization acquires an additional contribution $M=M_0+\delta M\theta(H-H^*)$ with
\begin{equation}\label{deltaM}
\delta M = -\frac{\lambda}{4\pi \xi L_z}\left(A_1+H\frac{\partial A_1}{\partial H}\right). 
\end{equation}
From Eq.~(\ref{constraint})  we have that the dependence on $A_0$ drops from Eq.~(\ref{EqA1}). Defining $a_1=A_1/(\xi H^*)$ and $h=H/H^*$, in the limit $\lambda\gg \xi$ we can cast the equation for $A_1$ in the form $2a_1=\epsilon\sqrt{1-1/(h-2a_1)^2}$, with  
\begin{equation}\label{Eq:Epsilon}
\epsilon=4\pi\times\frac{ev_F\rho_{\rm 2D}}{H^*}= 4\pi\times \frac{e v_F\rho_{\rm 2D}}{\eta H_{\rm c}},
\end{equation} 
For $H>H^*$ we choose the solution of $A_1$ that goes to zero at $H=H^*$. In a perturbative scenario, $\epsilon\ll 1$, $A_1$ is positive and its derivative is finite at $H^*$ and decreases with increasing $H$. It follows that the system develops an additional diamagnetic signal in the form of a negative jump at $H=H^*$ dominated by $H\partial A_1/\partial H$. More informations can be obtained by studying the susceptibility $\chi=\partial M/\partial H$ that, to lowest order in $\xi/\lambda$, reads
\begin{eqnarray}
\chi&=&\chi_0+\delta M(H^*)\delta(H-H^*)+\delta\chi,\\
\delta\chi&=&-\frac{\lambda}{4\pi L_z}\theta(h-1)\partial_h \left(a_1 +h\partial_h a_1\right),\label{deltachi}
\end{eqnarray}
with $\chi_0=\partial_HM_0$ the bare susceptibility. 
Formally, the susceptibility negatively diverges at $H=H^*$ as a result of the diamagnetic jump in $\delta M$. In Fig.~\ref{Figure}a) we plot $\delta\chi$ for different values of $\eta$. In the perturbative case $\epsilon=0.4$ ($\eta=0.5$) we see that $\delta\chi$ shows a positive peak at $H^*$, that is due to the strong increase of the current for $H>H^*$. Upon increasing $\epsilon$ the non-linear character of $\delta M$ fully manifests and $\delta\chi$ turns into a negative broad dip at $H=H^*$. 

We then conclude that in the overtilted regime $H>H^*$ the Majorana modes carry an Andreev current that adds to the bulk London current and participates to the Meissner screening. The resulting additional signal in the susceptibility shows a non-linear behaviour as a function of the applied field and an amplitude that scales as $\lambda/L_z$. In a system of linear size on order of micrometers a type II superconductor may show a penetration depth on order of tens on nanometers, thus making the amplitude of the additional signal accessible.  For thin slabs with $L_z\simeq \lambda$ the field penetrates the entire system and the critical threshold field shifts to $\eta H_{c2}$, that may well fall in the vortex phase, and the entire picture ceases to be valid.

\section{Curvature effects}
\label{Sec:curvature}
 
In a previous work \cite{Chirolli-SpinConn2018} we showed that the effective Hamiltonian Eq.~(\ref{HsurfA}) acquires an additional term on the surface of a sphere, that involves the spin connection of the Majorana modes. In spherical geometry it is possible to choose the self-consistent vector potential to have only non-zero azimuthal component ${\bf A}=\hat{\boldsymbol{\phi}}A_\phi(r)$, with $\nabla\cdot{\bf A}=0$.  Introducing Dirac matrices $\gamma^i\equiv(\gamma^0,\gamma^1,\gamma^2)=(i\alpha_x,\alpha_z,-\alpha_y)$, the Hamiltonian Eq.~(\ref{HsurfA}) gives rise to a Dirac equation on a flat Minkowski space $(\gamma^i+\gamma^0a^i)\partial_i\psi=0$ for the two-component Majorana spinor $\psi$, with $a^i=(e/m^* v_\Delta)\langle A^y\rangle \delta^i_{y}$, and the only non-zero component of $A^i$ is fixed to lay in the plane, orthogonal to the magnetic field. The Majorana equation on the surface of the sphere \cite{GonzalezPRL1992,ParentePRB2011} is then obtained by introducing the covariant derivative $D_\mu=\partial_\mu+\Gamma_\mu$, 
with $\Gamma_\mu$ the Majorana spin connection
\begin{equation}\label{Hdirac2D-A-S}
\left(\gamma^\mu+\gamma^0a^\mu\right)(\partial_\mu+\Gamma_\mu)\psi=0,
\end{equation}
where $\gamma^\mu=\gamma^ie_i^{~\mu}$,  $a^\mu=a^ie_i^{~\mu}$, and $e^i_{~\mu}\equiv\partial x^i/\partial x^\mu$. In addition, the Zeeman term acquires a finite component orthogonal to the surface in proximity of the poles of the sphere with the same matrix elements of the spin connection and it can be absorbed in the coupling $a^\mu$ (we refer to Ref.~[\onlinecite{Chirolli-SpinConn2018}] for details of the surface Majorana equation in presence of magnetic field and curvature). 

\subsection{Surface magnetic susceptibility}

We now calculate the surface magnetic susceptibility associated to the onset of a curvature-induced coupling between the external field and the Majorana cone via the Majorana spin connection. The surface Hamiltonian is
\begin{equation}
{\cal H}={\cal H}_0+{\cal H}_1\equiv \frac{\Delta}{k_FR}(\hat{H}_0+h\hat{H}_1)
\end{equation} 
with $h=H/H^*$ and ${\cal H}_0$ describing the unperturbed Majorana cone with eigenvalues $\epsilon^\alpha_{l}=\alpha (l+1/2)\Delta/(k_FR)$ associated to the eigenstates $|\Upsilon^\alpha_{lm}\rangle$, with $\alpha=\pm 1$, $l=1/2,3/2,\ldots$, and $|m|\leq l$ [\onlinecite{Chirolli-SpinConn2018}]. The perturbation $\hat{H}_1$ is specified by the only non-zero matrix elements 
\begin{equation}
\langle\Upsilon^+_{lm}|\hat{H}_1|\Upsilon^-_{l\pm 1,m}\rangle=\pm i\frac{\sqrt{(l+1/2\pm 1/2)^2-m^2}}{2(l+1/2\pm 1/2)}.
\end{equation}
For weak field we can calculate the correction to the total energy at second order in perturbation theory,
\begin{equation}\label{DeltaE}
\delta E=\sum_{l,l',m,\alpha,\beta}\frac{f(\epsilon^\alpha_{l})-f(\epsilon^\beta_{l'})}{\epsilon^\alpha_l-\epsilon^\beta_{l'}}
\left|\langle\Upsilon^\alpha_{lm}|{\cal H}^1|\Upsilon^\beta_{l'm}\rangle\right|^2,
\end{equation}
with $f(\epsilon)=(1+\exp(\epsilon/T))^{-1}$ the Fermi-Dirac distribution function. At zero temperature, summing over occupied states with $|m|\lesssim l$ and $l\lesssim l_{\rm max}\simeq k_FR$ we find $\delta E=-\frac{1}{3}\Delta H^2/(H^*)^2$. Thus, the bare surface response, as given by $\delta\chi_0=-\partial^2 \delta E/\partial H^2$, is paramagnetic and amounts to 
\begin{equation}\label{deltaChi0}
\delta\chi_0=\frac{2}{3}\frac{\Delta}{(H^*)^2}. 
\end{equation}
The susceptibility beyond perturbation theory can be calculated by diagonalisation of the full surface Hamiltonian. Introducing the eigenvalues $\epsilon_{n,m}$ of $\hat{H}_0+h\hat{H}_1$, we define the dimensionless susceptibility
\begin{equation}
\delta\bar\chi=-\frac{1}{k_FR}\frac{\partial^2}{\partial h^2}\sum_{n,m}\theta(-\epsilon_{n,m})\epsilon_{n,m},
\end{equation}
from which it follows the zero-field susceptibility $\delta\chi_0=\delta\bar\chi(h=0)\Delta/(H^*)^2$ Eq.~(\ref{deltaChi0}). Once again, the total current is the sum of the bulk contribution, that depends on the full self-consistent vector potential, and the Andreev contribution. The latter is given by the expectation value on the occupied states of $\hat{H}_1$ and, through the Hellmann-Feynman theorem, it can be written as
\begin{equation}
j_A(r)=-ev_F\rho_{\rm 2D}|\phi(r)|^2R^2\int_0^hdh'\delta\bar\chi(h'),
\end{equation}
where the Andreev bound state wavefunction can be approximated as $\phi(r)=\exp((r-R)/\xi)/\sqrt{\xi R^2}$ for $\xi/R\ll 1$. Maxwell equations allow us to set $A_1=\epsilon H^*\xi\int_0^hdh'\delta\bar{\chi}(h')/4$, with $\epsilon$ given in Eq.~(\ref{Eq:Epsilon}). Maxwell equations also fix a constraint between the external field $H$ and the vector potential on the surface. For simplicity we can choose the same constraint Eq.~(\ref{constraint}) that applies to the planar boundary (see Appendix \ref{App:ScreenedSusc}  for a discussion). This way, Eq.~(\ref{deltaM}) applies and it predicts an additional diamagnetic contribution $\delta M$ to the magnetization that arises due to the Andreev current. 

The diamagnetic behaviour is understood by the fact that the orbital current generated in response to a field tends to screen the applied field, thus producing a diamagnetic signal.  In particular the susceptibility for $\xi\ll \lambda$  reads
\begin{equation}\label{Eq:ScreenSusc}
4\pi \chi=-1+\frac{\lambda}{R}-\frac{\epsilon\lambda}{2R}\delta\bar\chi(H/H^*),
\end{equation}
up to a correction on order of $h\delta\bar\chi'$ that is negligible at small field.  
We find that $\delta\bar\chi$ has a very weak dependence on $h$, so that susceptibility does not depend on the field and 
can be approximated with its zero field value $\delta\bar{\chi}(0)=2/3$. We clearly see that Meissner screening results in an 
additional diamagnetic contribution that is due to the surface. The amplitude of the additional signal is controlled by 
$\epsilon$, that in turn is inversely proportional to $\eta=|m|/\mu$, and can be tuned through doping.

\section{Finite frequency response}
\label{Sec:freq-resp}

Finally, we address the response of a time-dependent curvature-induced surface coupling in the spherical geometry. We assume that a small external field probes the system at frequency $\omega$, $\delta H(t)=\delta H \cos(\omega t)$, and couples to the Majorana cone via the term ${\cal H}_1(t)=\delta H(t)\Xi_1$, with $\Xi_1\equiv \partial {\cal H}_1/\partial H=\hat{H}_1\Delta/(H^*k_FR)$. The finite frequency susceptibility in linear response theory is given by
\begin{equation}\label{DeltaChi}
\delta\chi(\omega)=-\sum_{l,l',m,\alpha,\beta}\frac{f(\epsilon^\alpha_{l})-f(\epsilon^\beta_{l'})}{\omega+\epsilon^\alpha_l-\epsilon^\beta_{l'}+i0^+}
\left|\langle\Upsilon^\alpha_{lm}|\Xi_1|\Upsilon^\beta_{l'm}\rangle\right|^2.
\end{equation}
The relation between Eq.~(\ref{DeltaChi}) and Eq.~(\ref{DeltaE},\ref{deltaChi0}) is manifest, in that $\delta\chi_0\equiv\delta\chi(\omega=0)=-\partial^2\delta E/\partial H^2$. Summing over states with $|m|\leq l$ and $l<l_{\rm max}\simeq k_FR$, for large $k_FR$ we find that the suscpetibility is approximated by
\begin{equation}\label{Eq:EnvChi}
\delta\chi(\omega)=\delta\chi_0\left(1-\frac{\omega}{2\Delta}\ln\left|\frac{\omega+2\Delta}{\omega-2\Delta}\right|+i\frac{\pi \omega}{4\Delta}\right).
\end{equation}
In Fig.~\ref{Figure}b) we show the real and imaginary part of the susceptibility as a function of the frequency, calculated by numerically evaluating Eq.~(\ref{DeltaChi}), together with the envelopes provided by Eq.~(\ref{Eq:EnvChi}). We see that the susceptibility follows the behaviour predicted by Eq.~(\ref{Eq:EnvChi}). In addition, the susceptibility shows fast oscillations that reflect the discrete nature of the spectrum, with level spacing $\Delta/k_FR$. We then conclude that by irradiation with a time-dependent field the Majorana cone responds with a finite signal and that a detection of Majorana modes becomes at reach on a curved geometry.

\section{Conclusions} 
\label{Sec:conclusions}

In this work we presented an analysis of the magnetic response of a 2D cone of Majorana modes in class DIII topological superconductors. 
An in-plane applied field gives rise to a tilting of the Majorana cone along the direction of the supercurrent, that results in an excess Andreev 
current beyond a threshold field $H^*$ and an associated additional diamagnetic magnetization. The extra signal has a non-linear dependence 
on the applied field and scales as the ratio between the penetration depth $\lambda$ and the sample thickness $L_z$, $\lambda/L_z$. For type 
II superconductors characterised by an appreciable ratio $\lambda/L_z$, the signal becomes detectable in magnetic susceptibility measurements. 
On a curved surface the tilting term acquires a geometric contribution involving the Majorana spin connection, that couples positive and negative 
energy states, and adds to a curvature-induced finite Zeeman term, allowing interband transitions. A finite susceptibility arises at finite frequency 
in response to a time-dependent magnetic field. Our findings open the way to detection of Majorana modes by magnetization measurements and 
via the application of time-dependent magnetic field.

\section{Acknowledgments} 

The authors acknowledge very useful discussions with F. de Juan, I. Grigorieva, and A. K. Geim. 
L.C. and F.G. acknowledge funding from from the European Union's Seventh Framework Program (FP7/2007-2013) 
through the ERC Advanced Grant NOVGRAPHENE (GA No. 290846), L.C. acknowledges funding from the Comunidad de 
Madrid through the grant MAD2D-CM, S2013/MIT-3007. F.G. acknowledges funding from the European Commission under 
the Graphene Falgship, contract CNECTICT-604391.

\appendix

\section{Screened susceptibility}
\label{App:ScreenedSusc}

Analogously to what we have done for the planar geometry we now estimate the screened susceptibility, 
taking into account the Meissner screening. First of all, we need to calculate the current. This is done by 
taking the matrix element of the current operator between the occupied eigenstates of the system. 
Neglecting the tilting of the bands, we write the Hamiltonian as
\begin{equation}
{\hat H}(y)=\frac{\Delta}{k_FR}(\hat{H}_0+y\hat{H}_1),
\end{equation} 
where $\langle\Psi^\alpha_{l,m}|\hat{H}_0|\Psi^\beta_{l',m'}\rangle=\alpha (l+1/2)\delta_{l,l'}\delta_{m,m'}\delta_{\alpha,\beta}$, for positive half-integer 
$l=1/2,3/2,\ldots$, $|m|\leq l$, $\alpha=\pm 1$, 
\begin{eqnarray}
\langle\Psi^+_{l,m}|\hat{H}_1|\Psi^-_{l',m'}\rangle&=&\delta_{m,m'}\left[i\frac{\sqrt{(l+1)^2-m^2}}{2(l+1)}\delta_{l',l+1}\right.\nonumber\\
&-&\left.i\frac{\sqrt{l^2-m^2}}{2l}\delta_{l',l-1}\right],
\end{eqnarray}
and $y=A(R)/(\lambda H^*)$, with $A(R)$ the value of the azimuthal fully self-consistent vector potential of the sphere surface.

The Andreev current is obtain by expressing the action of the current operator on the surface state wavefunction. Its generic expression
\begin{equation}
{\bf j}_A({\bf r})=\hat{\boldsymbol{\phi}}\frac{e}{m^*}\sum_{m,n}\psi^\dag_{m,n}({\bf r})\hat{p}_\phi\psi_{m,n}({\bf r})\theta(-\epsilon_{m,n}),
\end{equation}
can be assumed to acquire the form ${\bf j}_A({\bf r})=j_A(r)\sin\theta\hat{\boldsymbol{\phi}}$ and $j_A(r)$ can be expressed through the matrix $\hat{H}_1$
\begin{equation}
j_A(r)=|\phi(r)|^2\frac{\hbar^2}{2m^*}\frac{2\pi}{\Phi_0}\frac{1}{R}\sum_{n,m}\langle \Theta_{m,n}|\hat{H}_1|\Theta_{m,n}\rangle \theta(-\epsilon_{m,n}),
\end{equation}
with $|\Theta_{m,n}\rangle$ the full eigenstates of $\hat{H}(y)$ with eigenvalues $\epsilon_{m,n}$, $\theta(-\epsilon)$ the Fermi function 
at zero energy, and $\phi(r)$ the Andreev state wavefunction. We then introduce a dimensionless susceptibility
\begin{equation}
\delta\bar{\chi}(y)\equiv -\frac{1}{k_FR}\frac{\partial^2}{\partial y^2}\sum_{n,m}\epsilon_{m,n} \theta(-\epsilon_{m,n}).
\end{equation}
Through application of the Hellmann-Feynman theorem we have that
the Andreev current can be written as 
\begin{equation}
j_A(r)=-ev_F|\phi(r)|^2(k_FR)^2\int_0^ydy'\delta\bar{\chi}(y').
\end{equation}
The total current then reads
\begin{equation}
j_{\rm tot}(r)=-\frac{A(r)}{4\pi \lambda^2}+j_A(r),
\end{equation}
and has to satisfy the Maxwell equation
\begin{equation}
\frac{\partial^2 A}{\partial r^2}+\frac{2}{r}\frac{\partial A}{\partial r}-\frac{2A(r)}{r^2}=\frac{A(r)}{\lambda^2}-4\pi j_A(r).
\end{equation}
Clearly, the self-consistent vector potential assumes a form $A(r)=A_0(r)+A_1(r)$, with $A_0(r)=A_0~f(r)$, 
where $f(r)=-{\rm Im}[y_{-2}(ir/\lambda)] e^{-R/\lambda}>0$ and $y_\nu(z)$ is a spherical Bessel function of the second kind. 

To simplify the problem we assume the Andreev wavefunction to have the form 
\begin{equation}
|\phi(r)|^2=2u(r)/(\xi^2R), 
\end{equation}
with $u(r)=-{\rm Im}[y_{-2}(2ir/\xi)]e^{-2R/\xi}$. This way, $A_1(r)=A_1u(r)$ and we find
\begin{equation}
A_1=4\pi \frac{ev_Fk_F^2R}{2(1-\xi^2/(4\lambda^2))}\int_0^ydy'\delta\bar{\chi}(y').
\end{equation}
Matching  the vector potential and its derivative with their respective forms valid $r>R$ at the sphere boundary $r=R$ we find
\begin{equation}
A_0=\frac{3HR-2A_1(2u(R)+Ru'(R))}{2(2f(R)+Rf'(R))}.
\end{equation}
For $R\gg \lambda,\xi$ we can approximate $-{\rm Im}[y_{-2}(z)]\sim  e^z/(2z)$ that yields the relation
\begin{equation}\label{constraintSphere}
A_0=3HR-A_1.
\end{equation}
This expression provides a constraint that binds $A_0$ and $A_1$ on the surface of the sphere. As for the case of the planar 
geometry we find that the Andreev current reduces the amplitude of the screening term $A_0$.

Finally we need to calculate the Gibbs free energy. By integration by parts it can be written as
\begin{equation}
G=\int_{\rm surface}\frac{d{\bf S}}{8\pi V}\cdot\left(\frac{{\bf h}({\bf r})}{2}-{\bf H}\right)\times{\bf A}({\bf r}),
\end{equation}
and we can approximate it as $G=-HA(R)/(8\pi R)$, with the vector potential at the boundary given by 
$A(R)=(A_0\lambda+2A_1/\xi)/(2R)$. It then follows that the magnetization is
\begin{equation}
4\pi M=-H+\frac{\lambda}{4 R^2}\frac{\partial}{\partial H}\left[H(3HR-A_1+2A_1\xi/\lambda)\right]
\end{equation}
that for $\xi/\lambda\ll1$ can be simplified to
\begin{eqnarray}
4\pi M&=&-H+\frac{3\lambda}{2 R}H-\frac{\lambda}{4 R^2}\frac{\partial}{\partial H}(HA_1)\\
&=&-H\left(1-\frac{3\lambda}{2 R}\right)+4\pi \delta M
\end{eqnarray}
Writing $A_1=RH^*a_1$ with $a_1=\frac{\epsilon}{2}\int_0^ydy'\delta\bar{\chi}(y')$ 
we obtain the result
\begin{equation}
4\pi \delta M = -\frac{\lambda H^*}{4 R}\frac{\partial}{\partial h}(h a_1)
\end{equation}
At this point we define $y=3h/2-a_1/2$ and obtain the implicit equation $y=3h/2-(\epsilon/4)\int_0^ydy'\delta\bar{\chi}(y')$, 
that can be approximately solved to give $a_1=(\epsilon/2)\int_0^hdh'\delta\bar{\chi}(h')$. 
The susceptibility is then given by
\begin{equation}
4\pi \chi=-1+\frac{3\lambda}{2R}-\epsilon\frac{\lambda}{4R}(\delta\bar\chi(h)+h\delta\bar{\chi}'/2),
\end{equation}
that shows a diamagnetic correction to the susceptibility and agrees well with the result Eq.~(20) 
obtained with a simplified model.

\bibliography{Bibfile}

\end{document}